\documentclass[reprint,amsfonts, amssymb, amsmath,  showkeys,pra,nofootinbib , superscriptaddress, twocolumn,longbibliography,aps]{revtex4-2}
\usepackage{float}
\makeatletter

\let\newfloat\newfloat@ltx
\makeatother

\usepackage[english]{babel}

\usepackage[utf8]{inputenc}
\usepackage{graphics}
\usepackage{selinput}
\usepackage[normalem]{ulem}
\usepackage[shortlabels]{enumitem}

\usepackage{braket}
\usepackage{amsthm}
\usepackage{mathtools}
\usepackage{physics}
\usepackage{xcolor}
\usepackage{graphicx}
\usepackage[left=16mm,right=16mm,top=35mm,columnsep=15pt]{geometry} 
\usepackage{adjustbox}
\usepackage{placeins}
\usepackage[T1]{fontenc}
\usepackage{lipsum}
\usepackage{csquotes}
\usepackage{bm}
\usepackage{bbm}

\usepackage[linesnumbered,ruled,vlined]{algorithm2e}
\SetKwInput{kwInit}{Init}

\usepackage{changes}
\definechangesauthor[name=Ilya Safro, color=red]{is}

\usepackage{mathrsfs}

\def\w{\omega}

\newcommand{\fsnull}[1]{}
\newcommand{\old}[1]{}

\usepackage[makeroom]{cancel}
\usepackage[toc,page]{appendix}
\usepackage[colorlinks=true,citecolor=blue,linkcolor=magenta]{hyperref}

\usepackage{tikz}
\tikzset{every picture/.style=remember picture}

\newcommand{\AC}{\mathcal{A}}
\newcommand{\BC}{\mathcal{B}}

\newcommand{\FC}{\mathcal{F}}

\newcommand{\RC}{\mathcal{R}}
\newcommand{\SC}{\mathcal{S}}
\newcommand{\TC}{\mathcal{T}}

\renewcommand{\leq}{\leqslant}

\renewcommand{\vec}[1]{\boldsymbol{#1}}

\newcommand{\bs}{\textsf{BS}}

\def\be{\begin{equation}}
\def\ee{\end{equation}}
\def\bs{\begin{split}}
\def\e{\end{split}}
\def\ba{\begin{eqnarray}}
\def\bea{\begin{eqnarray}}

\def\tea{\end{eqnarray}}
\def\ea{\end{eqnarray}}
\def\eea{\end{eqnarray}}
\def\w{\omega}

\newcommand{\T}{\intercal}

\usepackage{dsfont}

\def\be{\begin{equation}}
\def\te{\end{equation}}
\def\ee{\end{equation}}
\def\ba{\begin{eqnarray}}
\def\bea{\begin{eqnarray}}

\def\tea{\end{eqnarray}}
\def\ea{\end{eqnarray}}
\def\eea{\end{eqnarray}}

\newcommand{\beq}{\begin{equation}}
\newcommand{\eeq}{\end{equation}}

\newcommand{\1}{\mathbbm{1}}

\newcommand{\sx}{\sigma_x}
\newcommand{\sy}{\sigma_y}
\newcommand{\sz}{\sigma_z}

\newcommand{\ord}[1]{\mathcal{O}\left( #1 \right)}

\makeatletter
\renewcommand\onecolumngrid{
\do@columngrid{one}{\@ne}
\def\set@footnotewidth{\onecolumngrid}
\def\footnoterule{\kern-6pt\hrule width 1.5in\kern6pt}
}
\renewcommand\twocolumngrid{
        \def\footnoterule{
        \dimen@\skip\footins\divide\dimen@\thr@@
        \kern-\dimen@\hrule width.5in\kern\dimen@}
        \do@columngrid{mlt}{\tw@}
}
\makeatother

\begin{document}

\preprint{APS/123-QED}

\title{Background cancellation for frequency-selective quantum sensing}
\author{Ricard Puig}
\email{ricard.puigivalls@epfl.ch}
\affiliation{Theoretical Division, Los Alamos National Laboratory, Los Alamos, NM 87545, USA}
\affiliation{Institute of Physics, Ecole Polytechnique F\'{e}d\'{e}rale de Lausanne (EPFL), CH-1015 Lausanne, Switzerland}
\affiliation{Centre for Quantum Science and Engineering, Ecole Polytechnique F\'{e}d\'{e}rale de Lausanne (EPFL), CH-1015 Lausanne, Switzerland}
\author{Nathan Constantinides}
\affiliation{Theoretical Division, Los Alamos National Laboratory, Los Alamos, NM 87545, USA}
\affiliation{University of Maryland, College Park, Maryland 20740, USA}
\author{Bharath Hebbe Madhusudhana}
\affiliation{MPA-Q, Los Alamos National Laboratory, Los Alamos, NM, USA}
\author{Daniel Bowring}
\affiliation{Fermi National Accelerator Laboratory, Batavia, Illinois 60510, USA}
\author{C. Huerta Alderete}
\affiliation{Information Sciences, CAI-3, Los Alamos National Laboratory, Los Alamos, NM, USA}
\author{Andrew T. Sornborger}
\email{sornborg@lanl.gov}
\affiliation{Information Sciences, CAI-3, Los Alamos National Laboratory, Los Alamos, NM, USA}

\date{\today}

\begin{abstract}
\noindent
A key challenge in quantum sensing is the detection of weak time dependent signals, particularly those that arise as specific frequency perturbations over a background field. Conventional methods usually demand complex dynamical control of the quantum sensor and heavy classical post-processing. We propose a quantum sensor that leverages time independent interactions and entanglement to function as a passive, tunable, thresholded frequency filter. By encoding the frequency selectivity and thresholding behavior directly into the dynamics, the sensor is responsive only to a target frequency of choice whose amplitude is above a threshold. This approach circumvents the need for complex control schemes and reduces the post-processing overhead. 

\end{abstract}
\maketitle

\paragraph*{Introduction}---
A core task in fundamental and biological sciences is detecting local perturbations in a field to identify rare events masked by a background signal or noise~\cite{frandsen2024new, kasevich1991atomic,shi2014sensing,zhao2012sensing,joshi2023method}. The ability to sensitively detect such low-probability events -- ranging from traces of rare molecules to small perturbations in a magnetic field -- can open new avenues in physics, biology, and engineering.

Of particular interest are rare events that satisfy specific criteria, such as time-dependent signals at a given frequency or particles with a defined energy, which may serve as markers for concrete processes or properties~\cite{cai2013diamond, pelliccione2016scanned,vcerenkov1937visible}.

Several techniques have been developed to mitigate unwanted effects from background fields on a sensor. For classical signals, interferometric configurations are often sufficient to cancel the background and isolate small perturbations~\cite{ronchi1964forty, aasi2015advanced}. This simple yet powerful approach underlies a wide range of experiments, including recent large-scale implementations such as LIGO~\cite{abbott2016observation}.

By contrast, detecting quantum signals requires the use of quantum sensors~\cite{degen2017quantum}. For a time-dependent Hamiltonian, $H(t)$, coupled to a variable signal, such sensors exploit carefully engineered states, measurements, and interactions to estimate properties of the underlying Hamiltonian~\cite{pang2017optimal, allen2025quantum}. When these states are subject to decoherence, several error-correction protocols have been developed~\cite{zhou2018achieving, gorecki2020optimal}. Another widely studied approach is Dynamical Decoupling~\cite{alvarez2011measuring}, which employs rapid control pulses to decouple a state from its environment. In addition, when a coherent background field is present, techniques based on precisely timed pulses can be used to isolate specific frequencies~\cite{kotler2011single, arrad2014increasing, slichter2013principles}. However, these pulse-based methods demand a high degree of experimental control, which can pose practical challenges and limit their robustness. Moreover, in the presence of a background field, they require independent characterization of the system with and without the perturbation, followed by post-processing to extract the desired signal.

The use of interactions in quantum sensing has recently gained significant traction~\cite{alderete2025nonlinear,montenegro2025quantum, abbasgholinejad2025optimally}. Interactions can enhance the sensitivity of a quantum sensor to the ultimate limit, even for equilibrium quantum metrology~\cite{puig2024dynamical}. They have also been employed in critical metrology~\cite{sarkar2025exponentially}, where sensors operating near critical points can approximate non-linear functions of time-dependent parameters and exhibit increased robustness to noise~\cite{alderete2025nonlinear}.

Here, we introduce a quantum sensor that harnesses controlled, time-independent, and local interactions to operate as a frequency filter. The protocol coherently cancels a background field, rendering the sensor responsive only to a local perturbation above background at a pre-selected frequency. This cancellation is effective even for random, shot-to-shot fluctuating backgrounds --- a unique feature of our technique. 

\paragraph*{Framework}---
Obtaining a simple response function that encodes information about a perturbation or signal $H_S(t)$ is, in general, a challenging task. In particular when minimal assumptions are made about $H_S(t)$. We consider the most general single qubit signal,
\begin{equation}
    H_S(t) =  \vec{s}(t) \cdot
    \vec{\sigma}\;,
\end{equation}
where $\vec{s}(t) = (s^{(x)}(t),s^{(y)}(t),s^{(z)}(t))$ defines the signal's direction and magnitude, and $\vec{\sigma} = (\sx,\sy,\sz)$ is the vector of Pauli operators. 

For such general fields, it is advantageous to design a protocol that captures key features of the signal without relying on detailed assumptions, ensuring applicability even when little is known about the field. Moreover, as highlighted above, it is desirable for the sensor to respond selectively, particularly to signals with specific frequency components. Determining the Fourier coefficients of $ \vec{s}(t)$ not only reveals the presence of signals at those frequencies but also enables, if needed, the reconstruction of the full vector $ \vec{s}(t)$.

Therefore, without loss of generality, we expand the signal vector $\vec{s}$ in the Fourier basis
\begin{align}\label{eq:s_fourier_decomposition}
     \vec{s}(t) = \sum_{\w} \vec{s}_{s_{\w}}\sin(\w t)+\vec{s}_{c_{\w}}\cos(\w t) \;,
\end{align}
where we choose the real basis since the signal must be Hermitian, $\vec{s}^*(t) = \vec{s}(t)$. 

Sensing a time-dependent signal is inherently non-trivial. Even in the simple case of a magnetic field $\vec{s}(t) = (0,0,s^{(z)}(t))$, probed with the state $\ket{+}$, the accumulated phase difference is $\int_0^T s^{(z)}(t) dt$, which may average to zero. In contrast, for a time-independent signal with constant strength $\theta$, the phase accumulates linearly as $\theta t$.
This contrast highlights the need for more elaborate strategies to extract information from time-dependent fields. The standard approach employs control pulses~\cite{hahn1950spin}, which approximately isolate the Fourier coefficients introduced in Eq.~\eqref{eq:s_fourier_decomposition}.
However, implementing such pulses demands a high level of experimental control that might not always be available. Moreover, in the presence of a background field $H_B(t) = \vec{b}(t) \cdot\vec{\sigma}$, these methods fail unless detailed knowledge of $H_B(t)$ is available to allow post-processing and background subtraction.

\begin{figure}
    \centering
    \includegraphics[width=0.4\textwidth] {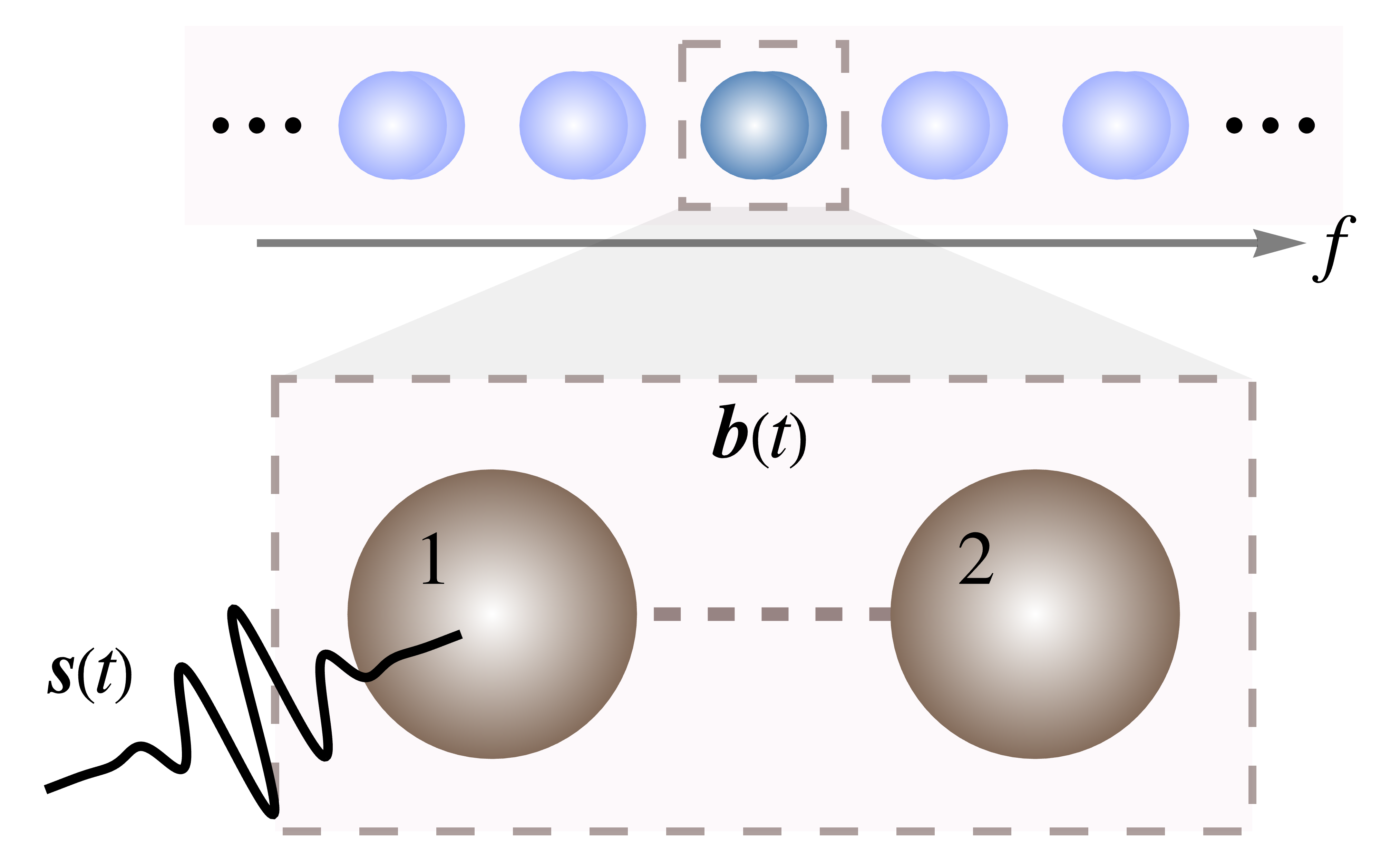}
    \caption{Two qubits are prepared in a Bell state. Qubit 1 couples to both the signal $ \vec{s}(t)$ and the background $\vec{b}(t)$, while qubit 2 couples only to the background. The joint evolution coherently cancels the background contribution, isolating the response to the signal.
    }
    \label{fig:schematic}
\end{figure}

\paragraph*{The two-qubit sensor}\label{sec:the_single_qubit_filter}--- 
We now introduce a sensing protocol designed to isolate the Fourier amplitudes of the signal while suppressing background contributions. The procedure consists of the three steps:
\begin{enumerate}
    \item Initialization: The system is prepared in the Bell state $\Psi_-$ where one qubit is subjected to the signal $H_S$ and both experience the background $H_B$, as depicted in Fig.~\ref{fig:schematic},
    \begin{align}\label{eq:initial_bell_state_appendix_general_main}
        \ket{\Psi_-} = \frac{1}{\sqrt{2}}(\ket{10}-\ket{01})\;.
    \end{align}
    \item Interaction: The  time-dependent Hamiltonians for the two qubits are
    \begin{align}\label{eq:H_1_main_general}
       H_1(t) =\frac{\w_0}{2} (\vec{n}_{\w_0}\cdot\vec{\sigma}) + [ \vec{s}(t)+\vec{b}(t)]\cdot\vec{\sigma}\,,
    \end{align}
    \begin{align}\label{eq:H_2_main_general}
       H_{2}(t) = \frac{\w_0}{2} (\vec{n}_{\w_0}\cdot\vec{\sigma}) + (\vec{b}(t)\cdot\vec{\sigma}) \,.
    \end{align}
    The qubits evolve for a time $T$, corresponding to one full period of $\vec{s}(t)$. The procedure is repeated for $\vec{n}_{\omega_0} = (\pm1,0,0)$, $(0,\pm1,0)$, and $(0,0,\pm1)$.  
    
    \item Measurement: The spins are measured along the direction of the control $(\vec{n}_{\w_0} \cdot\vec{\sigma})\otimes \1+\1\otimes(\vec{n}_{\w_0} \cdot\vec{\sigma})$. For weak signals, the probability of obtaining an even-parity outcome (e.g. $++,--$ for the $\sx$ measurements) is
    \begin{align}\label{eq:prob_approx}
        \overline{p}  \approx &\frac{T^2\|\vec{s}_{\w_0}\|_2^2}{12}\,,
    \end{align}
    where $\|\vec{s}_{\w_0}\|_2^2 = \|\vec{s}_{s_{\w_0}}\|_2^2 + \|\vec{s}_{c_{\w_0}}\|_2^2$ is the squared norm of the signal's Fourier components corresponding to the selected frequency $\w_0$. See Appendix~\ref{app:proof_general_protocol} for details.
\end{enumerate}

The response function in Eq.~\eqref{eq:prob_approx} shows that the sensor responds only when the signal contains a component at the target frequency and remains insensitive to the background field. 

\begin{figure}[t]
    \centering
    \includegraphics[width=\linewidth, trim={7 7.3 0 0},clip]{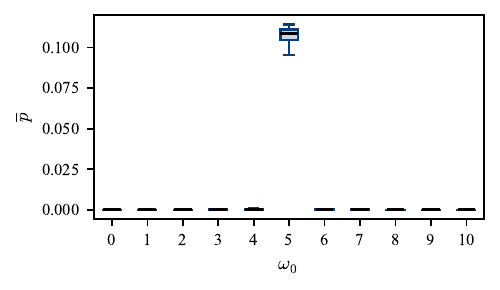}
    \caption{\textbf{Frequency selection of the protocol.} Box plot of the sensor response for different control frequencies $\omega_0$. Each data point corresponds to evolution over one signal period, with the background field randomized across 200 realizations. The background is a normalized random function with support on ten frequencies, while the signal is a single sinusoid at $\w = 10\pi$. 
    }
    \label{fig:response_only_when_necessary}
\end{figure}

Beyond estimating the amplitude, the protocol is also valuable for detection tasks~\cite{karsa2024quantum}. For instance, one might be interested in detecting whether the signal has support on a given frequency, a simpler task than amplitude estimation. In this case, the sensor is especially useful because, as shown in Appendix ~\ref{app:zero_signal}, an even-parity measurement outcome is strictly impossible when no signal component exists at the target frequency. The protocol can therefore serve as a single-shot detector. Figure~\ref{fig:response_only_when_necessary} illustrates this behavior. The exact response probability, given in Eq.~\eqref{eq:prob_approx}, is plotted as a function of the control frequency. The results show that the response vanishes for all frequencies where the corresponding amplitude of the signal is zero. 

Analytically, the error in the approximation in Eq.~\eqref{eq:prob_approx} is bounded by:
\begin{equation}\label{eq:error_in_response_fun}
    |r| \in \order{\left(\int_0^T dt\|\vec{s}(t)\|_2\right)^3}\, .
\end{equation}
 In other words, the estimator bias is at most cubic in the perturbation amplitude when the signal strength is comparable to the background. To verify this bound, we perform numerical simulations of the sensor's response.

\begin{figure}[t]
    \centering
    \includegraphics[width=\linewidth, trim={7 9 0 0},clip]{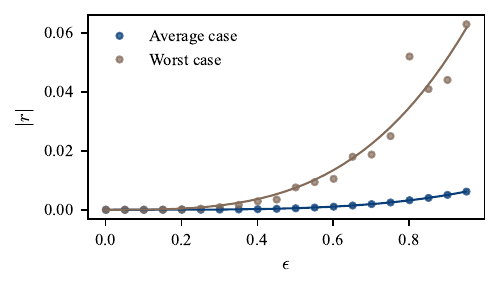}
    \caption{\textbf{Error of the response function.} Error $|r|$ in the estimated probability as a function of the signal strength $\epsilon = \|\vec{s}\|_2 = \|\vec{b}\|_2$, with random Fourier coefficients. The blue curve shows the average deviation between the estimator and the exact probability, while the brown  curve shows the maximum error observed. Polynomial regressions $f(\epsilon) = b\epsilon^3 + c \epsilon^4$ are fitted to both data sets, yielding $f_{\rm A}(\epsilon) =  9.78\times10^{-4}\,\epsilon^3 + 6.56\times10^{-3}\,\epsilon^4,\ R^2=0.9999 $ for the average case and $f_{\rm W}(\epsilon)=4.57\times10^{-2}\,\epsilon^3 + 2.74\times10^{-2}\,\epsilon^4,\ R^2=0.94681$ for the worse case. Errors are computed from 2000 randomly generated functions with fixed frequency support ($10$ and $7$ for $\vec{b}(t),\vec{s}(t)$ respectively).}
    \label{fig:polynomial_regression_of_error}
\end{figure}

Figure~\ref{fig:polynomial_regression_of_error} shows both the average and worst case (maximal) errors over several repetitions, where $\vec{b}(t)$ and $\vec{s}(t)$ are randomly sampled with bounded norm. For each repetition the two-norm of $\vec{s}(t)$ and $\vec{b}(t)$ are fixed and we compute the probability of obtaining the $\bra{++},\bra{--}$ states after evolving over a period $T$. A polynomial regression on the maximal and average errors confirms that the leading term scales as $\|\vec{s}(t)\|_2^3$. The small prefactors observed in the fit suggest a substantial constant improvement over the theoretical bound.

\paragraph*{Scaling}--- The previous section demonstrated the protocol's efficacy for a two-qubit system. Extending it to a multi-qubit system while preserving background cancellation and frequency filtering properties is, however, nontrivial. Here, we show how the protocol can be scaled to achieve the ultimate enhancement in the sensitivity, namely the Heisenberg scaling (HS)~\cite{toth2014quantum}. To obtain the HS one needs to compute the Classical Fisher Information (CFI), an attainable lower bound on the variance of any estimator. This quantity makes it possible to quantify how entanglement improves sensitivity, that is, while independent sensors exhibit a standard deviation that scales as $\Delta/\sqrt{N}$, entangled probes can reach $\Delta/N$, where $\Delta$ is the standard deviation of a single sensor. For an unknown parameter $\chi$, estimated from a probability distribution $p(\chi)$, the CFI is given by
\begin{align}\label{eq:cfi}
    \FC_C(\chi) = \frac{(\partial_\chi p(\chi))^2}{p(\chi)(1-p(\chi))}\,.
\end{align}

To illustrate this scalability, we consider a system of $2N$ qubits, partitioned into two subsystems of $N$ qubits each. Each subsystem forms a collective spin system with total spin $S = N/2$. The initial state is prepared as
 \begin{align}
     \ket{\psi} = \frac{1}{\sqrt{d}}\sum_{m_z=-S}^S(-1)^{S-m_z}\ket{S,m_z}\ket{S,-m_z}\,,
 \end{align}
where $\ket{S,m_z}$ are Dicke states~\cite{dicke1954coherence}, eigenstates of collective spin operators. Defining the collective spin operators as $S_i = \frac{1}{2}\sum_j^{n}\sigma_{i}^{(j)}$ for $i=x,y,z$, these states satisfy $S_z\ket{S, m_z} = m_z$ and $(S_x^2 + S_y^2 + S_z^2)\ket{S, m_z}= S(S+1)\ket{S, m_z}$. 

The two subsystems evolve independently under the Hamiltonians $\sum_{i=1}^{N}H_1^{(i)}(t)$ and $\sum_{i=1}^{N}H_2^{(i)}(t)$ defined in Eqs.~\eqref{eq:H_1_main_general} and~\eqref{eq:H_2_main_general}, respectively. After the evolution, we measure the collective spin operator $S_{\vec{n}_{\w_0}}\otimes\1 + \1\otimes S_{\vec{n}_{\w_0}}$. The presence of the signal is inferred from measurement outcomes where the two subsystems yield collective projections $m_{\vec{n}_{\w_0}}$ and $m'_{\vec{n}_{\w_0}}$, such that $m_{\vec{n}_{\w_0}} - m'_{\vec{n}_{\w_0}}\neq 0$.

In this case, $p(M)$ denotes the probability density per collective measurement outcome, rather than per pair of qubits. This probability scales quadratically with the number of qubits,
\begin{align}
    p(M) \approx N^2 \overline{p}\,,
\end{align}
where $\overline{p}$ is the probability from Eq.~\eqref{eq:prob_approx}. Substituting into Eq.~\eqref{eq:cfi}, the CFI scales as
\begin{align}
    \FC_C(\|\vec{s}_{{\w_0}}\|_2) \approx N^2 \,,
\end{align}
consistent with the Heisenberg-limit.

\paragraph*{Discussion}--- We have introduced a quantum sensor that functions as a passive, frequency-selective filter capable of eliminating the effects of background fields. By encoding both frequency filtering and background removal directly into the sensor's dynamics, we obtain a physical system that performs an {\em in situ} information processing task, effectively acting as a physical (quantum) computer~\cite{markovic2020physics,momeni2025training} that is sensitive to field perturbations only at a chosen frequency and above background. Owing to its simplicity, the sensor can be readily implemented in state-of-the-art platforms. For instance, bosonic alkali atoms such as $^{87}$Rb, $^{39}$K, $^{7}$Li or Cs trapped in optical tweezers, can be arranged in a ladder system, i.e., $2\times N$ array of individual atoms~\cite{Zhang2025_floating_phases, Liao2025_Rydberg_ladder_phase_diagram}. The $i$-th pair of qubits can be encoded in the two traps on the $i-$th step of the ladder. The qubits may use hyperfine states $\ket{0}\equiv \ket{F=1, m_F=0}$ and $\ket{1}\equiv \ket{F=1, m_F=-1}$ (for Cs, it would be $\ket{0}\equiv \ket{F=3, m_F=0}$ and $\ket{1}\equiv \ket{F=4, m_F=-1}$). The Hamiltonian $\frac{\omega_0}{2}\vec{n}_{\omega_0} \cdot \vec{\sigma}$ can be engineered via a detuned microwave drive with phase $\phi$, strength $\Omega$, and detuning $\delta$, yielding $\Omega \cos \phi \sigma_x+ \Omega \cos\phi \sigma_y + \delta \sigma_z$. To make qubit~2 insensitive to the signal, it can be placed physically distant by coherently transporting the optical tweezer trap post state preparation~\cite{Bluvstein2022CoherentTransport}. If the signal has a smaller spatial range than the background, i.e., the signal is more focused, this will ensure that qubit-2 is insensitive to the signal, but sensitive to the background. Another technique is to shelve qubit 2 in a magnetically insensitive state during phase acquisition, for example by mapping $\ket{F=2,m_F=-1}\mapsto\ket{F=2,m_F=0}$ via a two-photon transition. This will ensure that all the background during the entire sequence is corrected for, except for during the short time window of phase acquisition.

To understand the value of this type of sensor, we discuss this background cancellation approach in the context of searches for wave-like dark
matter (DM). Various theories of sub-eV, bosonic dark matter seek to explain the energy density
of the galactic DM halo through the coherent oscillation of classical DM waves
\cite{berlin2024, asztalos2006, caputo2021}. Via faint couplings to the standard model, these
theories predict the generation of anomalous electromagnetic waves that can be detected in the laboratory \cite{sushkov2023}. One class
of DM search experiments places photon detectors at the focal point of a broadband dish antenna
\cite{horns2013, knirck2018, liu2022, jeong2023, fan2025}. Similar techniques have also been proposed for searches for high-frequency gravitational waves
\cite{capdevilla2025}. In the setting considered here, the signal $\vec{s}(t)$ is localized at
the antenna's focal point, while various thermal and electronic backgrounds $\vec{b}(t)$ are broadly
distributed across the experimental apparatus. The velocity distribution
for the DM halo $v\simeq10^{-3}$  will result in the "smearing" of the antenna's focal point.
For the Broadband Reflector Experiment for Axion Detection (BREAD), this region is estimated to be $O(1)$~mm \cite{liu2022}. Notably, on-chip Bell states have been generated with larger
spatial separation than this
\cite{steffen2013, bhattacharjee2025}.

Beyond conventional sensing, the protocol can serve as a front-end module for quantum processors, providing pre-filtered quantum states for subsequent information processing~\cite{zhou2018achieving,gorecki2020optimal,allen2025quantum,puig2024variational,chin2025quantum}.

\paragraph*{Acknowledgments}---
We thank Martí Perarnau-Llobet, John Calsamiglia, Zoë Holmes and Marco Cerezo for insightful conversations. R.P. acknowledges the support of the SNF Quantum Flagship Replacement Scheme (Grant No. 215933). R.P. and N.C. were supported by the U.S. Department of Energy (DOE) through a quantum computing program sponsored by the Los Alamos National Laboratory (LANL) Information Science \& Technology Institute. C.H.A. and A.T.S. acknowledge initial support from the ASC Beyond Moore's Law project and subsequent research presented in this Letter was supported (sequentially) by the Laboratory Directed Research and Development program of Los Alamos National Laboratory under project numbers 20248110CT-IMS (C.H.A. and A.T.S.) and (C.H.A., B.H.M., and A.T.S.) 20251064ER. This paper is approved by Los Alamos National Laboratory for universal release under designation LA-UR-26-20119. This document was prepared using the resources of the Fermi National Accelerator Laboratory (Fermilab), a U.S. Department of Energy, Office of Science, Office of High Energy Physics HEP User Facility. Fermilab is managed by Fermi Forward Discovery Group, LLC, acting under Contract No. 89243024CSC000002.

\bibliography{local,quantum}

\onecolumngrid

\newpage

\appendix
\section*{Appendix: ``Background cancellation for frequency-selective quantum sensing''}

The appendices are organized as follows. Appendix~\ref{app:simple_freq_filter} introduces a simplified version of the frequency-filtering strategy to build intuition. Appendix~\ref{app:proof_general_protocol}, presents the full protocol and provides a detailed proof of its validity. We begin in Appendix~\ref{app:zero_signal} by showing that the state remains invariant in the absence of a signal, and in Appendix~\ref{app:freq_recovering} we demonstrate that, when the signal is present, the protocol approximately recovers its Fourier amplitudes. Appendix~\ref{app:error_bounds_general} derives an analytical bound on the estimator error. Finally, Appendix~\ref{app:Heisenberg_Scaling} extends the analysis to demonstrate  Heisenberg Scaling.

\section{Filtering a single amplitude of a frequency}\label{app:simple_freq_filter}
In this appendix, we show how the control interaction proportional to $\sigma_x$ can isolate the amplitude of a single frequency component of a classical field $f(t)$ coupled to $\sz$. Similar ideas have been explored in related contexts~\cite{schoelkopf2002qubits}.

We begin with the time-dependent Hamiltonian
\begin{align}
    H(t) = \w_0\sx/2 + f(t)\sz \;,
\end{align}
where $\w_0$ denotes the control frequency. The field $f(t)$ is expanded in a real Fourier basis over one period $T$ as, 

\begin{align}\label{eq:f_fourier_decomposition}
    f(t) = \sum_{\w} \left[f_{s_{\w}}\sin(\w t)+f_{c_{\w}}\cos(\w t) \right] \;,
\end{align}
with coefficients
\begin{equation}\label{eq:Fourier_coeff_f}
    f_{s_\omega} := \frac{2}{T}\int_{0}^{T} f(t) \sin(\omega t) dt \;, \qquad  f_{c_\omega} := \frac{2}{T}\int_{0}^{T} f(t) \cos(\omega t) dt \;.
\end{equation}
To analyze the dynamics, we move to the interaction picture with respect to the control term $\omega_{0} \sigma_x /2$, such that
\begin{align}\label{eq:interaction_picture}
    V(t) &= f(t) e^{-i\w_0 t\sx/2} \sz e^{i \w_0 t\sx/2} \nonumber \\
    &= f(t) \left[ \cos(\omega_{0} t) \sigma_{z} - \sin(\omega_{0} t) \sigma_{y}\right] \;.
\end{align}
Assuming the weak-coupling condition $\int_0^{T} |f(t)| dt \ll 1 $ we truncate the Dyson series~\cite{sakurai1995modern} at first order (defined with an overline), 
\begin{align}
    \overline{U(t)} &\approx \1 -i\int_0^T d\tau V(\tau)\;.
\end{align}
Using Eqs.~\eqref{eq:f_fourier_decomposition}-\eqref{eq:interaction_picture} and the orthogonality of the Fourier basis (e.g. $\int_0^T\sin(\w t)\sin(\w' t) = \delta_{\w,\w'} T/2$), the time-evolution operator becomes,
\begin{align}\label{eq:unitary_expansion_for_general_V}
    \overline{U(t)} &\approx \1 -i \int_0^T d\tau f(\tau) e^{-i\w_0 t\sx/2} \sz e^{i \w_0 t\sx/2} \nonumber \\
     &\approx \1 -i \sum_{\omega}\int_0^T d\tau \left[f_{s_{\w}}\sin(\w \tau)+f_{c_{\w}}\cos(\w \tau) \right] \left[ \cos(\omega_{0} \tau) \sigma_{z} - \sin(\omega_{0} \tau) \sigma_{y}\right] \nonumber \\
    &= \1 - \frac{i T}{2} \left( f_{c_{\omega_{0}}} \sigma_{z} - f_{s_{\omega_{0}}} \sigma_{y} \right) \;.
\end{align}
This expression shows that the Fourier coefficients $f_{c_{\omega_{0}}}$ and $f_{s_{\omega_{0}}}$ can be directly extracted. For instance, if the system is initialized in $\ket{0}$, the probability of finding it in $\ket{1}$ after an evolution time $T$ is determined solely by the $\sigma_{y}$ term at first order, yielding
\begin{align}
    p_{0\to 1} \approx \frac{T^2 f_{s_{\omega_0}}^2}{4}\,.
\end{align}

Thus, the single-qubit protocol isolates a specific Fourier component of the field. In the next appendix we combine this single-qubit frequency filter with a background-removal scheme in a two-qubit setting.

\section{Proof of the protocol}\label{app:proof_general_protocol}
In this appendix, we prove that the single qubit frequency filter recovers the desired amplitudes at the selected frequency while coherently canceling contributions from the background signal. For clarity, we restate the sensing protocol introduced in the main text:
\begin{enumerate}
    \item Initialization: The system is prepared in the Bell state $\Psi_-$ where one qubit is subjected to the signal $H_S$ and both qubits experience the background $H_B$,
    \begin{align}\label{eq:initial_bell_state_appendix}
        \ket{\Psi_-} = \frac{1}{\sqrt{2}}(\ket{10}-\ket{01})\;.
    \end{align}
    \item Interaction: The  time-dependent Hamiltonians for the two qubits are
    \begin{align}\label{eq:H_1_appendix}
       H_1(t) =\frac{\w_0}{2} (\vec{n}_{\w_0}\cdot\vec{\sigma}) +  [\vec{s}(t)+\vec{b}(t)]\cdot\vec{\sigma}\,,
    \end{align}
    \begin{align}\label{eq:H_2_appendix}
       H_{2}(t) = \frac{\w_0}{2} (\vec{n}_{\w_0}\cdot\vec{\sigma}) + (\vec{b}(t)\cdot\vec{\sigma}) \,,
    \end{align}
    where $\vec{s}(t)$ and $\vec{b}(t)$ denote the signal and background fields, respectively. The qubits evolve for a time $T$, corresponding to one full period of $\vec{s}(t)$. The procedure is repeated for $\vec{n}_{\omega_0} = (\pm1,0,0)$, $(0,\pm1,0)$, and $(0,0,\pm1)$ to access all Cartesian components.  
    
    \item Measurement: The spins are measured along the control direction $(\vec{n}_{\w_0} \cdot\vec{\sigma})\otimes\1+\1\otimes(\vec{n}_{\w_0} \cdot\vec{\sigma})$. For weak signals, the probability of obtaining an even-parity outcome (e.g. $++,--$ for the $\sx$ measurements) is
    \begin{align}\label{eq:prob_approx_appendix}
        \overline{p}  = &\frac{T^2\|\vec{s}_{\w_0}\|_2^2}{12} +\order{\left(\int_0^T |\epsilon_s(\tau)|d\tau\right)^3} \,,
    \end{align}
    where the subscript $i$ denotes either of the two even-parity outcomes in the chosen measurement basis, and $\|\vec{s}_{\w_0}\|_2^2 = \|\vec{s}_{s_{\w}}\|_2^2 + \|\vec{s}_{c_{\w}}\|_2^2$ is the squared norm of the signal's Fourier components corresponding to the selected frequency $\w_0$.
\end{enumerate}

\subsection{The protocol for a zero signal}\label{app:zero_signal}
We first show that the probe state remains unchanged when the signal $\vec{s}(t)$ is absent. Starting from the Bell state $\vert \Psi_{-} \rangle$, we consider the evolution dictated by Eqs.~\eqref{eq:H_1_appendix}-\eqref{eq:H_2_appendix} under the assumption that $\vec{s}(t) = 0\,\forall\, t\in[0,T]$. In this case, $H_1(t) = H_2(t)$, and the joint evolution of the two qubits can be written as
\begin{align}\label{eq:unitaries_s_eq_zero}
    U(t)\otimes U(t) \ket{\Psi_-} = -i U(t)\otimes U(t) \sy \ket{\Phi_+} = i U(t) \sy U^\T (t)\otimes \1 \ket{\Phi_+}
\end{align}
where we used $\1\otimes \sy \ket{\Phi_+} = i\ket{\Psi_-}$ to rewrite the Bell state in terms of $\ket{\Phi_+}$, the \textit{Ricochet property}~\cite{khatri2019quantum}, $A\otimes B \ket{\Phi_+} = A B^\T \otimes \1 \ket{\Phi_+}$, to move $U(t)$ through the tensor product, and the relation $\sy^\T = -\sy$ to simplify the overall sign.

The transpose of the time-evolution operator can be expressed explicitly using the time-ordering operator $U^\T (t)$ as
 \begin{align} \label{eq:U_T}
    U^{\T}(t) &= \left( \TC \exp \left[ -i\int_0^t d\tau~ \frac{\w_0}{2} \vec{n}_{\w_0}\cdot\vec{\sigma} + \vec{b}(\tau)\cdot\vec{\sigma}  \right] \right)^\T \nonumber\\
    &= \overline{\TC} \exp\left[ -i\int_0^t d\tau~ \frac{\w_0}{2} \overline{\vec{n}}_{\w_0}\cdot\vec{\sigma} + \overline{\vec{b}}(\tau)\cdot\vec{\sigma}  \right]
\end{align}
where $\overline{\vec{a}} = (a_x,-a_y,a_z)$ for $\vec{a} = (a_x,a_y,a_z)$, and $\overline{\TC}$ denotes the inverse ordering. For completeness, the time-ordering operator acts as
\begin{align}\label{eq:time_order_operator}
    \TC\{O(t_1)O(t_2)\} = 
    \begin{cases}
        O(t_1)O(t_2) \,{\rm ~if~}\, t_1>t_2\\
        O(t_2)O(t_1)\, {\rm ~if~}\, t_2>t_1\;.
    \end{cases}
\end{align}
Using the Pauli anti-commutation  relations and comparing term-by-term with the definition of  $U^{\dagger}(t)$, we obtain
\begin{align}\label{eq:sy_anticommute_U}
    \sy U^{\T}(t) &= \overline{\TC} \exp\left[ i\int_0^t d\tau \frac{\w_0}{2} {\vec{n}}_{\w_0}\cdot\vec{\sigma} + {\vec{b}}(\tau)\cdot\vec{\sigma}  \right] \sy = U^{\dagger}(t) \sigma_{y}\,.
\end{align}
Substituting this relation into Eq.~\eqref{eq:unitaries_s_eq_zero} gives

\begin{align}
    U(t)\otimes U(t) \ket{\Psi_-} &=i \sy\otimes \1 \ket{\Phi_+} = \ket{\Psi_-}
\end{align}
Hence, the state remains invariant when $\vec{s}(t)=0$, confirming that the protocol is insensitive to the background field alone.
In this regime, the system can therefore serve as a background-free quantum detector: any deviation from the initial state directly signals the presence of a local perturbation.

\subsection{The response function at first order for a nonzero signal}\label{app:freq_recovering}
We now consider the case of a weak signal. We are interested in the probability of measuring an even parity-outcome when the sensor is prepared in the Bell state $\ket{\Psi_{-}}$:
\begin{equation}
    p(i,i) =  \vert \langle ii \vert U_1(t)\otimes U_2(t)\ket{\Psi_-} \vert ^{2} \,
\end{equation}
where the even-parity projectors are taken in the basis aligned with the control axis: $\left\{ \ket{++}, \ket{--}\right\}$ for $\sigma_{x}$,  $\left\{ \ket{+i,+i}, \ket{-i,-i}\right\}$ for $\sigma_{y}$, and $\left\{ \ket{00}, \ket{11}\right\}$ for $\sigma_{z}$. 

In the weak-coupling regime, we truncate at first order (as in Appendix~\ref{app:simple_freq_filter}),
\begin{align}
     \overline{U_j(t)} =& \1 -i A_{j} (t) \qquad j=1,2\,,
\end{align}
where we have defined shorthand notation for the integrals
\begin{align}\label{eq:A_j}
    A_1(t) &= \int_0^t dt_1 [\mathcal{B}(t_1)+\mathcal{S}(t_1)], 
    &
    A_2(t) &= \int_0^t dt_1 \BC(t_1),
\end{align}
given in terms of the driven parts in the rotated (interaction) frame with respect to the control term $\frac{\w_0}{2} \vec{n}_{\w_0}\cdot\vec{\sigma}$,
\begin{equation}\label{eq:rotating_frame}
    \mathcal{B}(t) \equiv e^{-i t\w_0/2 \vec{n}_{\w_0}\cdot\vec{\sigma}} \vec{b}(t)\cdot\vec{\sigma}e^{+it\w_0/2 \vec{n}_{\w_0}\cdot\vec{\sigma}} , \qquad \mathcal{S}(t) \equiv e^{-it\w_0/2 \vec{n}_{\w_0}\cdot\vec{\sigma}} \vec{s}(t)\cdot\vec{\sigma}e^{+it\w_0/2 \vec{n}_{\w_0}\cdot\vec{\sigma}}.
\end{equation}
Using that $\overline{U_1(t)\otimes U_2(t)}\ket{\Psi_-} = \overline{U_1(t) U_2^\dagger(t) \otimes\1 }\ket{\Psi_-}$ (up to a global phase), it suffices to consider the first-order propagator $\overline{U_1(t)U_2^\dagger(t)}$. Thus, the first-order expansion reads
\begin{align}\label{eq:expansion_UsUdagger_first_order}
    \overline{U_1(T)U_2^\dagger(T)} =& \left( \1 - i A_{1} (T)\right)\left( \1 + i A_{2}(T)\right) \nonumber \\
    & \approx  \1  - i A_{1}(T) + i A_{2}(T) \nonumber \\
    & = -i\int_0^T d\tau \SC(\tau)\;,
\end{align}
i.e., the background cancels at this order. Using the adjoint action of a rotation generated by $\vec{n}_{\omega_0}\cdot\vec{\sigma}$,
\begin{align}\label{eq:adjoint_action} 
    \SC(\tau) = \vec{s}(\tau)\cdot\vec{\sigma}~\cos({\omega_0 \tau}) + (\vec{n}_{\omega_0} \times \vec{s}(\tau))\cdot \vec{\sigma}\sin({\omega_0 \tau}) + (\vec{n}_{\omega_0}\cdot\vec{\sigma})(\vec{n}_{\omega_0}\cdot\vec{s}(\tau))[1-\cos(\omega_0 \tau)]\:, \nonumber \\
\end{align}
we now simplify by taking $\vec{n}_{\w_0}$ along one of the cardinal directions $\{\hat{x},\hat{y},\hat{z}\}$. Each choice of control axis isolates the two components of $\vec{s}(t)$ perpendicular to that axis. Throughout we define the Fourier coefficients at $\omega_0$ by
\begin{equation}\label{eq:Fourier_coeff}
    \frac{2}{T} \int_0^T  s_i(\tau) \cos(\omega_0\tau) d\tau = s^{(i)}_{c_{\omega_0}}, 
    \qquad
    \frac{2}{T} \int_0^T  s_i(\tau) \sin(\omega_0\tau) d\tau = s^{(i)}_{s_{\omega_0}},
    \quad i\in\{x,y,z\}.
\end{equation}
\paragraph{ Control along $\vec{n}_{\omega_{0}} = \pm \hat{x}$.} For $\vec{n}_{\omega_0} = (\pm1,0, 0)$, the adjoint action in Eq.~\eqref{eq:adjoint_action} yields
\begin{align}
     \SC(\tau) = s_{x}(\tau) \sigma_{x} + [s_{y}(\tau) \cos(\omega_{0}\tau)\mp s_{z}(\tau) \sin(\omega_0\tau)]\sigma_{y} + [s_{z}(\tau) \cos(\omega_{0}\tau)\pm s_{y}(\tau) \sin(\omega_0\tau)]\sigma_{z}.
\end{align}
The measurable response lies in the $y$-$z$ plane, perpendicular to the control direction $\hat{x}$. Using
\begin{align}
    \sx\otimes\1 \ket{\Psi_-} = -\ket{\Phi_-}, \qquad
    \sy\otimes\1\ket{\Psi_-} = i \ket{\Phi_+}, \qquad
    \sz\otimes\1 \ket{\Psi_-} =&\ket{\Psi_+},
\end{align}
and $\ket{\pm \pm} =  \left( \ket{\Phi_{+}} \pm \ket{\Psi_{+}}\right)/\sqrt{2}$, one finds that,
\begin{align}
    \bra{++} \overline{U_1(T)\otimes U_2(T)}\ket{\Psi_-} &= \bra{++} \overline{U_1(T) U_2^\dagger(T) \otimes\1} \ket{\Psi_-} \nonumber\\
    &= -i\int_0^T d\tau \bra{++} \Big\{s_{x}(\tau) \sigma_{x} \otimes \1 + [s_{y}(\tau) \cos(\omega_{0}\tau)\mp s_{z}(\tau) \sin(\omega_0\tau)]\sigma_{y}\otimes\1 \nonumber\\
    & \quad + [s_{z}(\tau) \cos(\omega_{0}\tau)\pm s_{y}(\tau) \sin(\omega_0\tau)]\sigma_{z}\otimes\1\Big\} \ket{\Psi_{-}}
    \nonumber\\
    &= \frac{1}{\sqrt{2}} \left[ \int_0^T d\tau\left[ s_y(\tau)\cos(\w_0\tau) \mp s_z(\tau)\sin(\w_0\tau) \right] +i \int_0^T d\tau\left[ s_z(\tau)\cos(\w_0\tau) \pm s_y(\tau)\sin(\w_0\tau) \right] \right]\,.
\end{align}
Similarly, the amplitude for $\ket{--}$ flips the sign of the imaginary part. Thus, the first-order even-parity probability for control signal is along $\pm \hat{x}$ is 
\begin{align}\label{eq:probabilityxpm}
    \overline{p}_{\pm x}(+,+) = \overline{p}_{\pm x}(-,-) = \frac{T^2}{8}\left[\left( s^{(y)}_{c_{\w_0}} \mp s^{(z)}_{s_{\w_0}} \right)^2 + \left(s^{(z)}_{c_{\w_0}} \pm s^{(y)}_{s_{\w_0}}  \right)^2\right].
\end{align}
Averaging over the two control directions $\vec{n}_{\w_0} = (\pm 1,0,0)$ gives
\begin{align}\label{eq:probabilityx}
    \overline{p}_{x} &= \frac{1}{2} \Big( \overline{p}_{+ x}(\pm, \pm)+ \overline{p}_{-x}(\pm, \pm)\Big) \nonumber \\
    &= \frac{T^2}{8} \left[(s^{(z)}_{c_{\w_0}})^2 + (s^{(z)}_{s_{\w_0}})^2 +  (s^{(y)}_{s_{\w_0}})^2 + (s^{(y)}_{c_{\w_0}})^2 \right],
\end{align}
which depends only on the $y$ and $z$ components of $\vec{s}(t)$ at $\omega_0$ (as expected from symmetry). Different postprocessing choices can isolate different linear combinations if desired.
\medskip\\
 
\paragraph{ Control along $\vec{n}_{\omega_{0}} = \pm \hat{y}$.} 
For $\vec{n}_{\omega_0} = (0,\pm1,0)$, the adjoint action in Eq.~\eqref{eq:adjoint_action} yields
\begin{align}
     \SC(\tau) = \left[s_x(\tau)\cos(\w_0 \tau)+ s_z(\tau)\sin(\w_0\tau)\right]\sx +  s_y(\tau)\sy +  \left[ s_z(\tau)\cos(\w_0\tau) - s_x(\tau)\sin(\w_0\tau) \right]\sz .\nonumber \\
\end{align}
As before, the response lies in the plane perpendicular to the control axis, now the $x$-$z$ plane. In this case, measuring both qubits in the $\sigma_{y}$ basis suffices, since the relevant even-parity Bell superpositions map onto $\ket{\pm i, \pm i} =  \left( \ket{\Phi_{-}} \pm i \ket{\Psi_{+}}\right)/\sqrt{2}$. Then, evaluating the transition amplitude,
\begin{align}
    \bra{+i,+i} \overline{U_1(T)\otimes U_2(T)}\ket{\Psi_-}  &= \frac{1}{\sqrt{2}} \left[ \int_0^Td\tau\left[s_x(\tau)\cos(\w_0 \tau) \pm  s_z(\tau)\sin(\w_0\tau)\right] + i \int_0^Td\tau\left[ s_z(\tau)\cos(\w_0\tau) \mp  s_x(\tau)\sin(\w_0\tau) \right] \right]. \nonumber \\
\end{align}

The first-order even-parity probability for control along $\pm\hat y$ then reads
\begin{align}
    \overline{p}_{\pm y}(+ i, + i) = \overline{p}_{\pm y}(- i, - i) =  \frac{T^2}{8}\left[\left( s^{(x)}_{c_{\w_0}} \mp s^{(z)}_{s_{\w_0}} \right)^2 + \left(s^{(z)}_{c_{\w_0}} \pm s^{(x)}_{s_{\w_0}}  \right)^2 \right].
\end{align}
Averaging over the two control orientations gives
\begin{align}\label{eq:probabilityy}
    \overline{p}_{y} =& \frac{1}{2} \Big( \overline{p}_{+ y}(\pm i, \pm i ) + \overline{p}_{- y}(\pm i,\pm i)\Big) \nonumber \\
    =& \frac{T^2}{8}\left[(s^{(x)}_{c_{\w_0}})^2 + (s^{(x)}_{s_{\w_0}})^2 +  (s^{(z)}_{s_{\w_0}})^2 + (s^{(z)}_{c_{\w_0}})^2 \right]
\end{align}
This result depends only on the components of $\vec s(t)$ perpendicular to the $\hat y$ axis, completing the symmetry with the $\pm\hat x$ case.\\
\medskip
\paragraph{ Control along $\vec{n}_{\omega_{0}} = \pm \hat{z
}$.} Finally, for $\vec{n}_{\w_0}= (0,0,\pm1)$  the adjoint action in Eq.~\eqref{eq:adjoint_action} gives
\begin{align}
     \SC = \left[s_x(\tau)\cos(\w_0 \tau)\mp s_y(\tau)\sin(\w_0\tau)\right]\sx +  \left[ s_y(\tau)\cos(\w_0\tau) \pm s_x(\tau)\sin(\w_0\tau) \right]\sy +  s_z(\tau)\sz
\end{align}
Thus, the response lies in the $x$–$y$ plane, perpendicular to the control direction $\hat{z}$. In this case, measuring both qubits in the $\sigma_z$ basis suffices, since the even-parity Bell combinations map onto $\ket{00} =  \left( \ket{\Phi_{+}} \pm \Phi_{-}\right)/\sqrt{2}$ and  $\ket{11} =  \left( \ket{\Phi_{+}} - \Phi_{-}\right)/\sqrt{2}$. Hence, evaluating the amplitude for $\ket{00}$,
\begin{align}
     \bra{00} \overline{U_1(T)\otimes U_2(T)}\ket{\Psi_-} = \frac{1}{\sqrt{2}} \left[\int_0^T d\tau \left[s_x(\tau)\cos(\w_0 t)\mp s_y(\tau)\sin(\w_0t)\right] + i\int_0^T d\tau\left[ s_y(\tau)\cos(\w_0\tau) \pm s_x(\tau)\sin(\w_0\tau) \right] \right]
\end{align}
By direct analogy with Eq.~\eqref{eq:probabilityxpm}, the corresponding even-parity probability for control along $\pm\hat{z}$ is
\begin{align}
    \overline{p}_{\pm z}(0,0) =  \overline{p}_{\pm z}(1,1) = \frac{T^{2}}{8} \left[ \left( s^{(x)}_{c_{\w_0}} \mp s^{(y)}_{s_{\w_0}} \right)^2 + \left(s^{(y)}_{c_{\w_0}} \pm s^{(x)}_{s_{\w_0}}  \right)^2\right]
\end{align}
Averaging over both control directions yields
\begin{align}\label{eq:probabilityz}
    \overline{p}_{z} =& \frac{1}{2} \Big( \overline{p}_{+ z}(i,i) + \overline{p}_{- z}(i,i)\Big) \nonumber \\
    =& \frac{T^2}{8} \left[(s^{(x)}_{c_{\w_0}})^2 + (s^{(x)}_{s_{\w_0}})^2 +  (s^{(y)}_{s_{\w_0}})^2 + (s^{(y)}_{c_{\w_0}})^2 \right]
\end{align}
with $i=0,1$.

Combining the three averaged results $\overline{p}_x$, $\overline{p}_y$, and $\overline{p}_z$ from Eqs.~\eqref{eq:probabilityx}, \eqref{eq:probabilityy}, and \eqref{eq:probabilityz}, the total even-parity probability averaged over the six control orientations is
\begin{align}\label{eq:first_order_final_prob}
    \overline{p} &= \frac{1}{3} \sum_{\mu=x,y,z} \overline{p}_{\mu} = \frac{T^{2}}{12} \sum_{\mu=x,y,z} \left[ (s_{c_0}^{(\mu)} )^{2} + (s_{s_0}^{(\mu)} )^{2} \right].
\end{align}
This expression shows that, to first order, the average even-parity probability directly encodes the total Fourier weight of the signal at the selected frequency $\omega_0$, completing the proof of Eq.~\eqref{eq:prob_approx_appendix}.

\subsection{Bound on the error} \label{app:error_bounds_general}
So far we have computed first-order probabilities. It is important to quantify the truncation error and, in particular, the bias of the resulting estimator. To do so, we start from the once-recursed Dyson form for each unitary,
\begin{align}
     U_j(t) =& \1 -i A_{j} (t) - Q_{j}(t)  \qquad j=1,2\,,
\end{align}
where we have defined shorthand notation for the integrals $A_{j}(t)$ in Eq. \eqref{eq:A_j} and 
\begin{align} 
    Q_1(t) &= \int_0^t dt_1 \int_0^{t_1} dt_2 [\mathcal{B}(t_1)+\mathcal{S}(t_1)][\mathcal{B}(t_2)+\mathcal{S}(t_2)] U_1(t_2), \nonumber \\
    Q_2(t) &= \int_0^t dt_1 \int_0^{t_1} dt_2 \mathcal{B}(t_1)\mathcal{B}(t_2) U_2(t_2).
\end{align}

In what follows, for simplicity we will omit the explicit temporal dependence when no confusion arises. Expanding the tensor product, \begin{align}
    U_1(t)\otimes U_2(t)
    = (\1 - iA_1 - Q_1) \otimes(\1 - iA_2 - Q_2),
\end{align}
we define the first–order truncation (denoted with an overline)
\begin{align}
    \overline{U_1(t)\otimes U_2(t)}
    := \1\otimes\1
    - i\bigl(A_1\otimes\1 + \1\otimes A_2\bigr)\,,
\end{align}
We use this to bound the distance between the exact product unitary and the first-order approximation. Collecting all higher-order terms in the reminder, 
\begin{align}
    \mathcal{R}(t) := U_1(t)\otimes U_2(t) -  \overline{U_1(t)\otimes U_2(t)},
\end{align}
and applying the triangle inequality,
\begin{align}
    \| \mathcal{R} \|_{\infty} \leq \|Q_1\|_\infty + \|Q_2\|_\infty
       + \|A_1\|_\infty \|Q_2\|_\infty
       + \|Q_1\|_\infty \|A_2\|_\infty
       + \|A_1\|_\infty \|A_2\|_\infty
       + \|Q_1\|_\infty \|Q_2\|_\infty 
\end{align}

To bound each term by using (i) the sub-multiplicativity for operator norms $\|A\otimes B\|_\infty = \|A\|_{\infty}\|B\|_{\infty}$; (ii) the triangle inequality for integrals $\Big\|\int M(t)\,dt\Big\|_\infty \le \int \|M(t)\|_\infty dt$; (iii) the conjugation invariance under $e^{\mp i\frac{\w_0 t}{2}\vec{n}_{\w_0}\cdot\vec{\sigma}}$, so that $\|\SC(t)\|_\infty = \|s(t)\|_\infty \leq \| s(t) \|_2 $ and $\|\BC(t)\|_\infty = \|b(t)\|_\infty \leq \|b(t) \|_2 $;   (iv) unitary invariance of the operator norm, $\|U(t)\|_\infty=1$; and (v) that for arbitrary functions $\left\vert\int_0^t dt_1 \int_0^{t_1} dt_2 f(t_1)g(t_2)\right\vert \leq \left(\int_0^t dt_1 |f(t_1)| \right) \left( \int_0^{t} dt_1 |g(t_{1})| \right)$. Then, we can define the following quantities to bound the reminder,
\begin{align}
    \| A_1 \|_{\infty} &\leq I_{b} + I_{s}, 
    & \| Q_1 \|_{\infty} &\leq (I_{b} + I_s)^{2}, \nonumber \\
    \| A_2 \|_{\infty} &\leq I_{b},
    & \| Q_2\|_{\infty} &\leq I_{b}^{2},
\end{align}
where
\begin{equation}
    I_s=\int_0^T \|s(t)\|_{2}~ dt,\qquad I_b=\int_0^T \|b(t)\|_{2}~dt.
\end{equation}
Combining these results gives 
\begin{align}
    \| \mathcal{R} \|_{\infty} &\leq (I_b+I_s)^2
       +I_b^2
       + (I_b+I_s)\,I_b^2
       +  (I_b+I_s)^2\,I_b
       + (I_b+I_s)\,I_b+ (I_b+I_s)^2 I_b^2.
\end{align}
In the weak-drive regime where $I_b+I_s\le 1$, the dominant scaling is quadratic:
\begin{align}
    \|\mathcal{R}\|_\infty
    = \bigl[(I_b+I_s)^2 + I_b^2 + (I_b+I_s)I_b\bigr]
      + \mathcal{O}\!\bigl((I_b+I_s)^3\bigr)
    = \mathcal{O}\!\bigl((I_b+I_s)^2\bigr).
    \label{eq:Rbound_BB_scaling}
\end{align}
Thus, all contributions beyond the explicit first–order and cross terms 
are suppressed quadratically in the integrated field strengths. 
Therefore, any overlap computed with the exact dynamics can be written as the approximate overlap plus a controlled error, this is
 \begin{align}
     \bra{\phi}U_1(T)\otimes U_2(T)\ket{\psi} =& \bra{\phi} \overline{U_1(T)\otimes U_2(T)}\ket{\psi} + \tilde{r} \label{eq:exact_response_with_eps}
 \end{align}
 with the amplitude-level reminder bounded by
 \begin{align} \label{eq:amplitude_error}
      \tilde{r} &= \vert \bra{\phi} (U_1(T)\otimes U_2(T) - \overline{U_1(T)\otimes U_2(T)}) \ket{\psi} \nonumber \\
      &\leq \| U_1(T)\otimes U_2(T) - \overline{U_1(T)\otimes U_2(T)} \|_{\infty} \nonumber  = \|\RC(T)\|_{\infty}\\
      &\leq (I_b+I_s)^2
       +I_b^2
       + (I_b+I_s)\,I_b^2
       +  (I_b+I_s)^2\,I_b
       + (I_b+I_s)\,I_b+ (I_b+I_s)^2 I_b^2.
 \end{align}

\subsection{The final response function}\label{app:final_response_single_w_error_general}
Finally, we combine the results from the two previous sections to obtain the full response function and its associated error. The total probability can be written as
\begin{align}
    p  = \overline{p} + r
\end{align}
where $\overline{p}$ is the first-order even-parity probability averaged over orientations, Eq. \eqref{eq:first_order_final_prob}, and $r: =p- \overline{p}$ collects all higher-order corrections (i.e. the error probability). Using the amplitude-level remainder $\tilde{r}$ bounded in Eq.~\eqref{eq:amplitude_error}, we can estimate
\begin{align}
   r =  |p-\overline{p}| = \Big|| \bra{\phi} \overline{U_1(T)\otimes U_2(T)}\ket{\psi} + \tilde{r} |^2 - \overline{p} \Big|  \leq 2 \sqrt{\overline{p}}|\tilde{r}| + |\tilde{r}|^2 \;.
\end{align}
Since $\sqrt{\overline{p}} \sim I_{s}$ and $|\tilde r|=\mathcal{O}((I_b+I_s)^2)$ in the weak-coupling regime (with $I_b\sim I_s$ for simplicity), the dominant contribution scales as
\begin{equation}
    r \lesssim \vert \tilde{r} \vert  I_{s} \in \order{ \left[ \int_{0}^{T} \|s (t) \|_{2}~ dt\right]^{3}}
\end{equation}

Thus, while the leading response in Eq.~\eqref{eq:first_order_final_prob} scales quadratically with the integrated signal strength, the residual bias in the probability is suppressed by an additional power of $I_s$, confirming that the first-order approximation faithfully captures the sensor’s response in the weak-drive regime.

\section{Heisenberg scaling}\label{app:Heisenberg_Scaling}
Entangling $N$ probes to coherently accumulate signal is a standard route to enhanced precision in quantum sensing, improving the error from the standard-quantum-limit scaling $1/\sqrt{N}$ to the Heisenberg scaling $1/N$ \cite{giovannetti2006quantum,pezze2018quantum}. In our case, the protocol is both frequency-selective and background-canceling, so
extending it to many particles requires checking that (i) background suppression survives collective control, and (ii)
the first-order, frequency-filtered response grows linearly in $N$ at the amplitude level (and hence quadratically in the
probability).

We consider a composite system of size $2N$, consisting of two $N$-qubit registers. The first register is subject to both signal and background fields, while the other is exposed only to the background. Each register is restricted to its symmetric subspace of total spin $S=N/2$ described by the collective operators
\begin{equation}\label{eq:vec_id_sym}
    S_{\mu} = \frac{1}{2} \sum_{i}^{N} \sigma_{\mu}^{(i)}, \qquad \mu=x,y,z. 
\end{equation}
In this subspace, we define the ``vectorized identity'' state as 

\begin{align}
    \ket{\1_{\rm sym}} = \frac{1}{\sqrt{d}}\sum_{m_z = -S}^S\ket{S, m_z}\ket{S, m_z} \qquad d= 2S+1 = N+1
\end{align}
where $S$ denotes the total spin and $m_z$ represents the value of the $S_z$. As in the two-qubit case, we apply $\sy^{\otimes N}$ to the second register, leading to the initial state
\begin{align}\label{eq:initial_multiple}
    \ket{\psi} = \frac{1}{\sqrt{d}}\sum_{m_z = -S}^S(-1)^{S-m_z}\ket{S, m_z}\ket{S, -m_z}\,.
\end{align}

Because the evolution generated by the background field is collective and identical on both registers, the same arguments as in Appendix \ref{app:proof_general_protocol} shows that the background cancels exactly. Hence we can focus on the first order response to the weak signal acting on the first register only. 

For identical signal coupling on all spins, the interaction in the rotating frame for a single spin is $S(t)$ as given in Eq. \eqref{eq:rotating_frame}. Summing over the $N$ spins in the {\em sensing} register gives $ \sum_{i=1}^{N} \mathcal{S}^{(i)}(t) = 2 \vec{s}(t)\cdot \vec{S}$ with $\vec{S} = (S_{x}, S_{y}, S_{z})$. The {\em background-only} register contributes an identical collective term involving $\vec{b}(t)\cdot \vec{S}$, which cancels at first order when we consider the propagator $U_{N,1}(T)U_{N,2}^\dagger(T)$, exactly as in Eq. \eqref{eq:expansion_UsUdagger_first_order}. Thus, the first-order propagator in the interaction picture for the two resister system is
\begin{align}\label{eq:expansion_UsUdagger_first_order_hs}
   \overline{U_{N,1}(T)U_{N,2}^\dagger(T)} =& -i2   \int_0^T d\tau \Big[\vec{s}(\tau)\cdot\vec{S}~\cos({\omega_0 \tau}) + (\vec{n}_{\omega_0} \times \vec{s}(\tau))\cdot \vec{S}\sin({\omega_0 \tau}) \nonumber\\
   & \qquad +(\vec{n}_{\omega_0}\cdot\vec{S})(\vec{n}_{\omega_0}\cdot\vec{s}(\tau))[1-\cos(\omega_0 \tau)]\Big] \:,
\end{align}
which is the same adjoint action as Eq. \eqref{eq:adjoint_action}, but with Pauli operators $\sigma$ replaced by collective spins $\vec{S}$.

As before, we specialize to control directions $\vec{n}_{\omega_0}$ aligned with the Cartesian axes. For concreteness, we take $\vec{n}_{\omega_0} = (1,0,0)$; the other cases follow by symmetry. In this case,
\begin{align}
     \sum_{i=1}^{N} \mathcal{S}^{(i)}(\tau) = s_{x}(\tau) S_{x} + [s_{y}(\tau) \cos(\omega_{0}\tau)- s_{z}(\tau) \sin(\omega_0\tau)]S_{y} + [s_{z}(\tau) \cos(\omega_{0}\tau)+ s_{y}(\tau) \sin(\omega_0\tau)]S_{z}\,.
\end{align}
From this expression we can clearly see that, using Eq.~\eqref{eq:Fourier_coeff}, the evolution operator $\overline{U_{N,1}(T)U_{N,2}^\dagger(T)}$ when $\vec{n}_{\omega_0} = (1,0,0)$ has the following form
\begin{align}\label{eq:multiple_qubits_x}
    -i2\int_0^T \sum_{i=1}^{N} \mathcal{S}^{(i)}(\tau) d\tau = -i \int_0^T d\tau s_{x}(\tau) S_{x} -i T[s_{c_{\w_0}}^{(y)} - s_{s_{\w_0}}^{(z)} ] S_y -i T[s_{c_{\w_0}}^{(z)} + s_{s_{\w_0}}^{(y)} ] S_z\,.
\end{align}
Thus, as in the single-qubit case, the response lies in the plane perpendicular to the control direction. We now work in the $S_{x}$ eigenbasis and consider measurement $S_x\otimes\1 + \1 \otimes S_x$. To precisely analyze the measurement we decompose it into its eigenprojectors. That is
\begin{align}
    S_x\otimes\1 + \1\otimes S_x = \sum_{m_x,m_x'}(m_x-m'_x)\ketbra{m_x}\otimes\ketbra{m_x'}\,.
\end{align}

From this expansion we see that the possible measurement outcomes (i.e. the projective spaces with different eigenvalues) are
\begin{align}
    \Pi_x(c) = \sum_{m}\ketbra{m}\otimes \ketbra{m-c}\,,
\end{align}
that corresponds to $m_x-m_x' = c$. In other words $ S_x\otimes\1 + \1\otimes S_x = \sum_{c=-S_x}^{S_x} c \,\Pi_x(c)$. 

Once we have established the projectors, we can compute the probability of obtaining a certain outcome. We can readily compute the probability of obtaining each of the $\Pi_x(c)$, that we will call $p_x(c)$. By recalling the form of the initial state $\ket{\psi}$ in Eq.~\eqref{eq:initial_multiple}, and the truncated unitary at first order along the $x$ direction shown in Eq.~\eqref{eq:multiple_qubits_x}, the value of $p_x(c)$ for any $c\neq 0$ will be
\begin{align}
    p_{+x}(c) =T^2 \expval{\AC_{+x}\otimes \1 \Pi_x(c)\AC_{+x} \otimes \1 }{\psi}
\end{align}
where we denoted $\AC_{+x} = [s_{c_{\w_0}}^{(y)} - s_{s_{\w_0}}^{(z)} ] S_y +T[s_{c_{\w_0}}^{(z)} + s_{s_{\w_0}}^{(y)} ] S_z$. By writing $S_y$ and $S_z$ in terms of the $S_x$-ladder operators $S_{\pm}^{(x)}$ we can rewrite $p_x(c)$ as 
\begin{align}
     p_{+x}(c) = &\frac{T^2}{2}\left|[s_{c_{\w_0}}^{(z)} + s_{s_{\w_0}}^{(y)} ]-i[s_{c_{\w_0}}^{(y)} - s_{s_{\w_0}}^{(z)} ]\right|^2 \expval{(S_{+}^{(x)}\otimes \1 )\, \Pi_x(c)\,  (S_{+}^{(x)} \otimes \1 )}{\psi} \\
     &+ \frac{T^2}{2}\left|[s_{c_{\w_0}}^{(z)} + s_{s_{\w_0}}^{(y)} ]+i[s_{c_{\w_0}}^{(y)} - s_{s_{\w_0}}^{(z)} ]\right|^2 \expval{(S_{-}^{(x)}\otimes \1 )\, \Pi_x(c)\,  (S_{-}^{(x)} \otimes \1 )}{\psi}\,,
\end{align}
an expression that we can further simplify using that $\sigma_y^{\otimes N}\ket{m_x} \sim \ket{-m_x}$ and recalling $\overline{p}_{+x}(+,+)$ from Eq.~\eqref{eq:probabilityxpm}
\begin{align}
     p_{+x}(c) = &\frac{T^2}{2d}\left|[s_{c_{\w_0}}^{(z)} + s_{s_{\w_0}}^{(y)} ]-i[s_{c_{\w_0}}^{(y)} - s_{s_{\w_0}}^{(z)} ]\right|^2 \left[2S(S+1) -  \frac{c}{2}\right] = \frac{4}{d}\left[2S(S+1) -  \frac{c}{2}\right] \overline{p}_{+x}(+,+)\,.
\end{align}

Therefore, if we consider all the possible $c\neq 0$, denoted by $\overline{p}_{M_x}$, the corresponding total probability is
\begin{align}
    \overline{p}_{M_x} = \frac{4\overline{p}_{+x}(+,+)}{d}\sum_{c=-S|\, c\neq 0}^S \left[2S(S+1) -  \frac{c}{2}\right] \overline{p}_{+x}(+,+) = \frac{S^2(S+1)}{2S+1}\overline{p}_{+x}(+,+)\, ,
\end{align}
where we used that $d = 2S+1$. Since $S=N/2$ we have that $S^2(S+1)/(2S+1)\sim \ord{N^2}$, so the collective probability along $x$ scales as
\begin{equation}
    \overline{p}_{M_x} \approx \mathcal{O}(N^2) \,\overline{p}_x.
\end{equation}
By symmetry, the same holds for control along $y$ and $z$:
\begin{equation}
    \overline{p}_{M_y} \approx \mathcal{O}(N^2)\,\overline{p}_y,\qquad
    \overline{p}_{M_z} \approx \mathcal{O}(N^2)\,\overline{p}_z.
\end{equation}
Averaging over the three control axes yields a total success probability
\begin{align}
    p(M) 
    &\approx \frac{1}{3} \left( \overline{p}_{M_x}+\overline{p}_{M_y}+\overline{p}_{M_z}\right)
      \sim N^2 \,\overline{p}\,,
\end{align}
where $\overline{p}$ is the single-pair probability in Eq.~\eqref{eq:first_order_final_prob}. Thus, the collective protocol amplifies the single-pair response by a factor $\sim N^2$ at the probability level.
\paragraph*{Classical Fisher information and Heisenberg scaling}---
To assess metrological performance, we consider the classical Fisher information (CFI) associated with the parameter $\|\vec{s}_{\omega_0}\|_2$, which controls the strength of the Fourier component at frequency $\omega_0$. For a binary outcome with success probability $P\equiv p(M)$, the CFI is
\begin{align}
    \mathcal{F}_C\big(\|\vec{s}_{\omega_0}\|_2\big) 
    = \frac{\big(\partial_{\|\vec{s}_{\omega_0}\|_2}P\big)^2}{P(1-P)}.
\end{align}
From the single-pair result in Eq.~\eqref{eq:first_order_final_prob}, $\overline{p}\propto \|\vec{s}_{\omega_0}\|_2^2$, so in the collective case $P \sim N^2 \|\vec{s}_{\omega_0}\|_2^2$ for weak signals ($P\ll1$). Hence
\begin{equation}
    \partial_{\|\vec{s}_{\omega_0}\|_2} P \sim N^2 \|\vec{s}_{\omega_0}\|_2,
\end{equation}
and
\begin{align}
    \mathcal{F}_C\big(\|\vec{s}_{\omega_0}\|_2\big) 
    \sim \frac{N^4 \|\vec{s}_{\omega_0}\|_2^2}{N^2 \|\vec{s}_{\omega_0}\|_2^2}
    \sim N^2.
\end{align}
Thus, the CFI scales as $N^2$, corresponding to Heisenberg-limited sensitivity. Since the Cramér–Rao bound implies
\begin{align}
    \Delta \|\vec{s}_{\omega_0}\|_2 \gtrsim \frac{1}{\sqrt{\mathcal{F}_C\big(\|\vec{s}_{\omega_0}\|_2\big)}}
    \sim \frac{1}{N}\,,
\end{align}
the multi-qubit extension of our protocol achieves Heisenberg scaling while preserving both frequency selectivity and background cancellation.

\end{document}